\documentclass[aps,floatfix,nofootinbib,showpacs,twocolumn,superscriptaddress]{revtex4}
\usepackage{graphicx}
\usepackage{dcolumn}
\usepackage{amsmath}
\usepackage{longtable}

\newcommand{\sfrac}[2]{\mbox{\footnotesize $\displaystyle \frac{#1}{#2}$}}
\newcommand{\bea}{\begin{eqnarray}}
\newcommand{\eea}{\end{eqnarray}}
\newcommand{\pvecabs}{\left|\vec p\right|}

\newcommand{\trD}{{\rm tr_D}}
\newcommand{\atan}{{\rm atan}}

\newcommand{\Real}{\mbox{\sf Re}}

\usepackage{color}
\definecolor{purple}{rgb}{0.5,0,0.5}
\definecolor{blue}{rgb}{0.0,0,0.9}
\usepackage[colorlinks=true, pdfstartview=FitV, linkcolor=purple, citecolor= purple, urlcolor=blue]{hyperref}

\begin{document}
\title{
Cold quarks in medium: an equation of state
}

\author{Thomas Kl\"ahn}
\affiliation{Physics Division, Argonne National Laboratory, Argonne,
Illinois 60439, USA}
\affiliation{Instytut Fizyki Teoretycznej, Uniwersytet Wroc\l{}awski,
pl.\ M.\ Borna 9, 50-204 Wroc\l{}aw, Poland}

\author{Craig D.~Roberts}
\affiliation{Physics Division, Argonne National Laboratory, Argonne, Illinois 60439, USA}
\affiliation{Department of Physics, Peking University, Beijing 100871, China}

\author{Lei Chang}
\affiliation{Institute of Applied Physics and Computational Mathematics, Beijing 100094, China}

\author{Huan Chen}
\affiliation{
Institute of High Energy Physics, Chinese Academy of Science, Beijing 100049, China}

\author{Yu-Xin Liu}
\affiliation{Department of Physics, Peking University, Beijing 100871, China}
\affiliation{State Key Laboratory of Nuclear Physics and Technology, Peking University, Beijing 100871, China}
\affiliation{Center of Theoretical Nuclear Physics, National Laboratory of Heavy Ion Accelerator, Lanzhou 730000, China}

\begin{abstract}
We derive a compact, semi-algebraic expression for the cold quark matter equation of state (EoS) in a covariant model that exhibits coincident deconfinement and chiral symmetry restoring transitions in-medium.
Along the way we obtain algebraic expressions for: the number- and scalar-density distributions in both the confining Nambu and deconfined Wigner phases; and the vacuum-pressure difference between these phases, which defines a bag constant.
%
The confining interaction materially alters the distribution functions from those of a Fermi gas and consequently has a significant impact on the model's thermodynamic properties, which is apparent in the EoS.
\end{abstract}
\maketitle

A reliable equation of state (EoS) for cold quark matter would be extremely valuable in modern astrophysics.  As emphasised, e.g., by Ref.\,\cite{Alford:2006vz}, the identification of a neutron star with a quark matter core depends upon it.  The problem is broader.  An almost complete absence of experimental constraints at densities above nuclear saturation entails that little is truly known about the EoS of any form of dense matter beyond that point \cite{Klahn:2009rw}.

In attempting to predict the properties and astrophysical signals of cold quark matter, all we have currently are models.  Owing, amongst other things, to the so-called fermion sign problem, the numerical simulation of lattice-QCD will not supply this need in the foreseeable future \cite{Alford:2009ar}.  Two classes of models are widely used in this application: bag- and Nambu--Jona-Lasinio-models.  As usually formulated, bag-like models possess a form of confinement but cannot describe dynamical chiral symmetry breaking (DCSB), whilst Nambu--Jona-Lasinio (NJL) models express DCSB but not confinement.  (NB.\ The static potential measured in quenched lattice-QCD is not related in any known way to the question of light-quark confinement, which can be connected with the analytic properties of QCD's Schwinger functions \protect\cite{Roberts:2007ji}.)

Dyson-Schwinger equations (DSEs) provide a continuum approach to QCD that can simultaneously address both confinement and DCSB.  They have been applied with success to hadron physics in-vacuum; e.g., \cite{Roberts:2007jh,Cloet:2008re,Chang:2009zb}, and to QCD at nonzero chemical-potential and temperature \cite{Roberts:2000aa}.  In the context of cold quark matter, the stability and properties of various Cooper-paired phases was recently explored via a truncation of the gap equation \cite{Nickel:2006vf} but that study did not provide information on the EoS.

An attempt to calculate the EoS was made in Ref.\,\cite{Zong:2008sm}.  However, in adopting a meromorphic model for the quark propagator and failing to consider the role played by the pressure difference between the Wigner and Nambu phases, this study could not describe a first-order chiral symmetry restoring transition.  As exhibited in Ref.\,\cite{Bender:1997jf} and elucidated in Ref.\,\cite{Chen:2008zr}, coincident, first-order chiral and deconfinement transitions are the natural results.

We will develop an EoS for a Poincar\'e covariant model \cite{Munczek:1983dx}, whose gap equation's solution does not admit a meromorphic parametrisation, and which exhibits coincident, first-order deconfinement and chiral symmetry restoring transitions.  A numerical analysis that exposes aspects of this model's EoS is described in Ref.\,\cite{Blaschke:1997bj}.  However, we obtain analytic results.  In particular, for the quark number- and scalar-density distributions, $f_1(\left|\vec p\right|;\mu)$ and $f_2(\left|\vec p\right|;\mu)$, which will themselves be useful, e.g., in astrophysics explorations.  Furthermore, these formulae are helpful in elucidating novel possibilities for the behaviour of fermions subjected to an interaction that supports a confining Nambu-Goldstone phase.

The in-medium, dressed-quark propagator is
\cite{Rusnak:1995ex}
\begin{eqnarray}
\nonumber
\lefteqn{
S(p;\mu)^{-1} = i \vec{\gamma}\cdot \vec{p} \, A(p^2,p\cdot u) }\\
%
&+&  i \gamma_4(p_4+i\mu) \, C(p^2,p\cdot u) + B(p^2,p\cdot u)  \,,
%
\label{sinvp}
\end{eqnarray}
where $u=(\vec{0},i\mu)$, with $\mu$ the quark chemical potential, and, in our Euclidean metric: $\{\gamma_\rho,\gamma_\sigma\} = 2\delta_{\rho\sigma}$; $\gamma_\rho^\dagger = \gamma_\rho$.  (NB.\ We employ an ultraviolet-finite model and hence no discussion of regularisation or renormalisation is necessary.)  The propagator is obtained from the gap equation
\begin{eqnarray}
S(p;\mu)^{-1} & = & i \vec{\gamma}\cdot \vec{p} + i\gamma_4(p_4+i\mu) +m + \Sigma(p;\mu)\,,\\
\nonumber
\Sigma(p;\mu ) &=& \int\frac{d^4 q}{(2\pi)^4}\, g^2(\mu) D_{\rho\sigma}(p-q;\mu) \\
& & \times \frac{\lambda^a}{2}\gamma_\rho S(q;\mu) \Gamma^a_\sigma(q,p;\mu) , \label{gensigma}
\end{eqnarray}
where $m$ is the bare mass, $D_{\rho\sigma}(k;\mu)$ is the dressed-gluon propagator and $\Gamma^a_\sigma(q,p;\mu)$ is the dressed-quark-gluon vertex.  

We specify the model through the choices \cite{Munczek:1983dx}
\begin{equation}
\label{Gk}
g^2 D_{\rho\sigma}(k)  = \left( \delta_{\rho\sigma} - \frac{k_\rho k_\sigma}{k^2} \right) 4\pi^4 \eta^2 \delta^4(k)\,,
\end{equation}
with $\eta$ a mass-scale parameter, and $\Gamma^a_\sigma(q,p)=\sfrac{1}{2}\lambda^a\gamma_\sigma$.  This vertex defines a rainbow gap equation, which is the leading-order in a systematic, symmetry-preserving DSE truncation scheme \cite{Munczek:1994zz,Bender:1996bb}.  The infrared enhancement exhibited by Eq.\,(\ref{Gk}) provides for confinement
and DCSB \cite{Roberts:2007jh}, and the model is super-asymptotically-free because the interaction strength vanishes for nonzero relative momentum.  In practice, the model has many features in common with a class of renormalisation-group-improved effective-interactions; and its distinctive momentum-dependence works to advantage in reducing integral- to algebraic-equations that preserve the character of the original.  It has been used widely with success; e.g., in exploring the impact of dressing the quark-gluon vertex \cite{Bhagwat:2004hn,Watson:2004kd,Matevosyan:2006bk} and in illuminating general, exact results connected with the $U_A(1)$ anomaly \cite{Bhagwat:2007ha}.  

In the chiral limit, the nonperturbative, chiral symmetry preserving solution of the model gap equation is $\hat A(p^2,p\cdot u)= \hat C(p^2,p\cdot u)$,
\begin{equation}
\hat C(p^2,p\cdot u) =\sfrac{1}{2}
\left(
	1+\sqrt{1+\sfrac{2\eta^2}{\tilde p^2}}
\right)\,,\; \hat B(p^2,p\cdot u)  \equiv  0\,,
\label{CABW}
%
%
\end{equation}
where $\tilde p^2 = p^2 + 2 p\cdot u + u^2$.  It describes a phase in which chiral symmetry is realised in the Wigner-Weyl mode and the quark is not confined.  For the analysis which follows, it is important to note that $\hat C(p^2,p\cdot u)$ possesses a branch point at $2\eta^2+\vec{p}\,^2+p_4^2-\mu^2 = 2 p_4 \mu=0$.  For $\mu \neq 0$, it occurs at $p_4=0$, $\vec{p}\,^2+2 \eta^2=\mu^2$.  Hence, the branch point plays a role when \mbox{$\mu^2 < 2 \eta^2$}.

The gap equation also has a confining solution, in which chiral symmetry is dynamically broken; viz., for $m=0$, $A(p^2,p\cdot u)  = C(p^2,p\cdot u)$,
\begin{eqnarray}
\label{CNG}
C(p^2,p\cdot u) & = &
\left\{
\begin{array}{ll}
2 & \Real(\tilde p^2) < \frac{\eta^2}{4}\\
\frac{1}{2}
\left(
	1+\sqrt{1+\frac{2\eta^2}{\tilde p^2}}
\right) & \mbox{otherwise,}
\end{array}
\right.\\
\label{BNG}
B(p^2,p\cdot u) &=&
\left\{
\begin{array}{ll}
\sqrt{\eta^2 - 4 \tilde p^2} & \Real(\tilde p^2) < \frac{\eta^2}{4}\\
0 & \mbox{otherwise.}
\end{array}
\right.
\end{eqnarray}
It describes a phase in which chiral symmetry is realised in the Nambu-Goldstone mode.  Confinement is signalled by a square-root branch point at $\tilde p^2 = \eta^2/4$, associated with the scalar piece of the self energy.  For $\mu \neq 0$, it occurs at $p_4=0$, $\vec{p}\,^2=\mu^2+\eta^2/4$.

Equations~(\ref{CABW}) and (\ref{CNG}), (\ref{BNG}) are all one needs in order to obtain this model's zero-temperature EoS.  NB.\ We ignore quark Cooper pairing herein.  Diquark condensates have previously been considered within this model \protect\cite{Bloch:1999vk}. We will extend our analysis to that case in the future.

We express the single-quark number density
\begin{eqnarray}
\label{nqmu}
n_q(\mu)& =& 2 \int\frac{d^3 p}{(2\pi)^3} \, f_1(|\vec{p}|;\mu)\,, \\
\label{nqmuf1}
f_1(|\vec{p}|;\mu) &= &\frac{1}{4\pi} \int_{-\infty}^\infty \! {\rm d}p_4 \, {\rm tr}_{\rm D} [-\gamma_4 S(p;\mu)]\,,
\end{eqnarray}
where the trace is over spinor indices alone.  In the Wigner phase, using Eqs.\,(\ref{CABW}), one obtains
\begin{equation}
\label{f1W}
f_1^W(\pvecabs;\mu) =
\left\{\begin{array}{ll}
1\,, & \vec p\,^2<\mu^2-2\eta^2 \\[1.5ex]
{\rm f}^W(\pvecabs;\mu) \,, & \mu^2-2\eta^2<\vec p\,^2<\mu^2\\[1.5ex]
0\,, & \mu^2<\vec p\,^2
\end{array} \right.
\end{equation}
where, with $\Delta= [\mu^2 - |\vec{p}\,|^2]/[2\eta^2]$,
%
\begin{equation}
\label{eq:f1WItwo}
{\rm f}^W(\pvecabs;\mu) =
1 +\sfrac{2}{\pi} \left[ \sqrt{ \Delta^2 (1-\Delta^2) } - \arccos \Delta\right].
%
\end{equation}

\begin{figure}[t]
\vspace*{-5ex}

\centerline{\includegraphics[width=0.33\textwidth,angle=-90,clip]{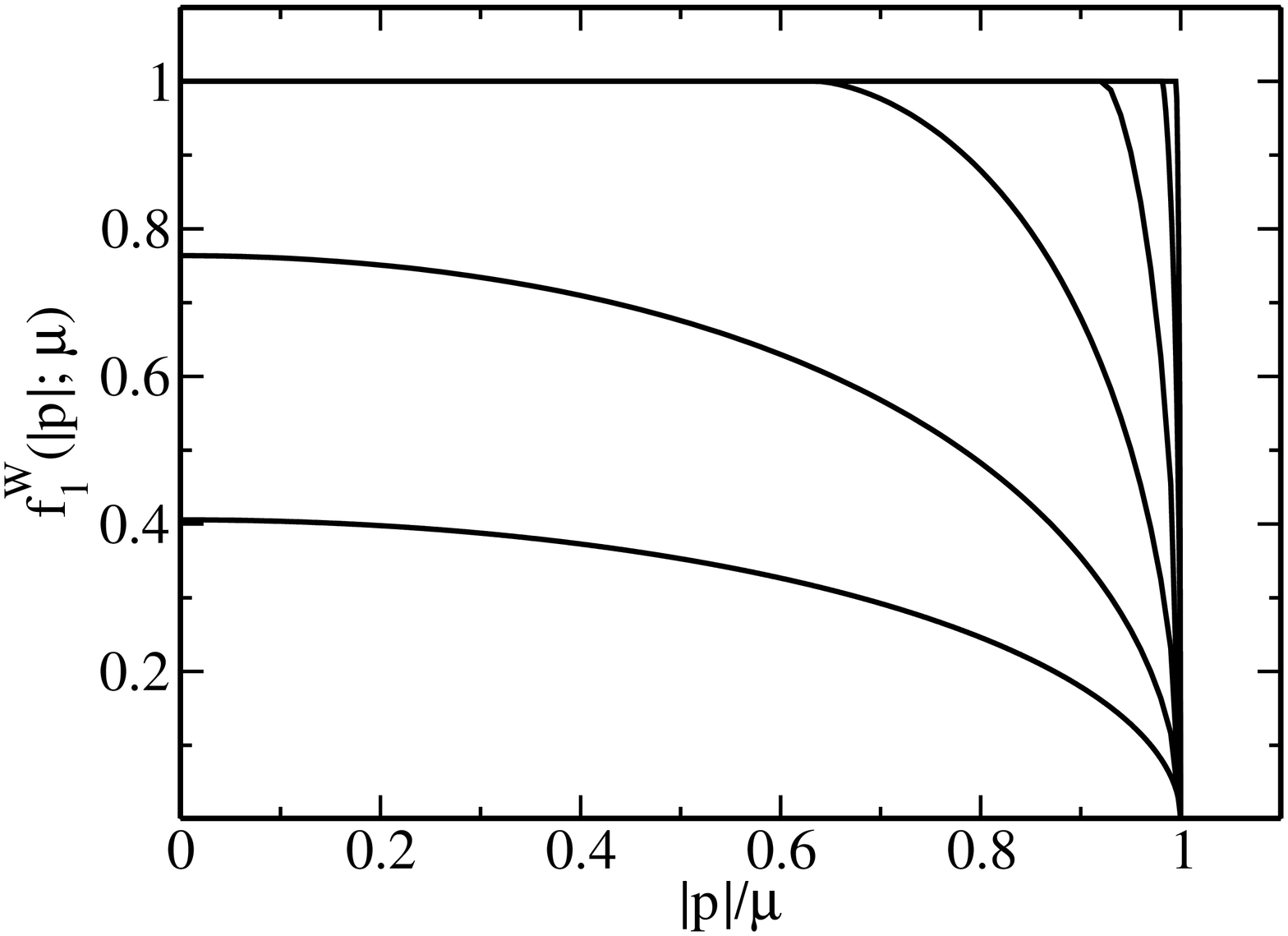}}
\vspace*{-2ex}

\centerline{\includegraphics[width=0.33\textwidth,angle=-90,clip]{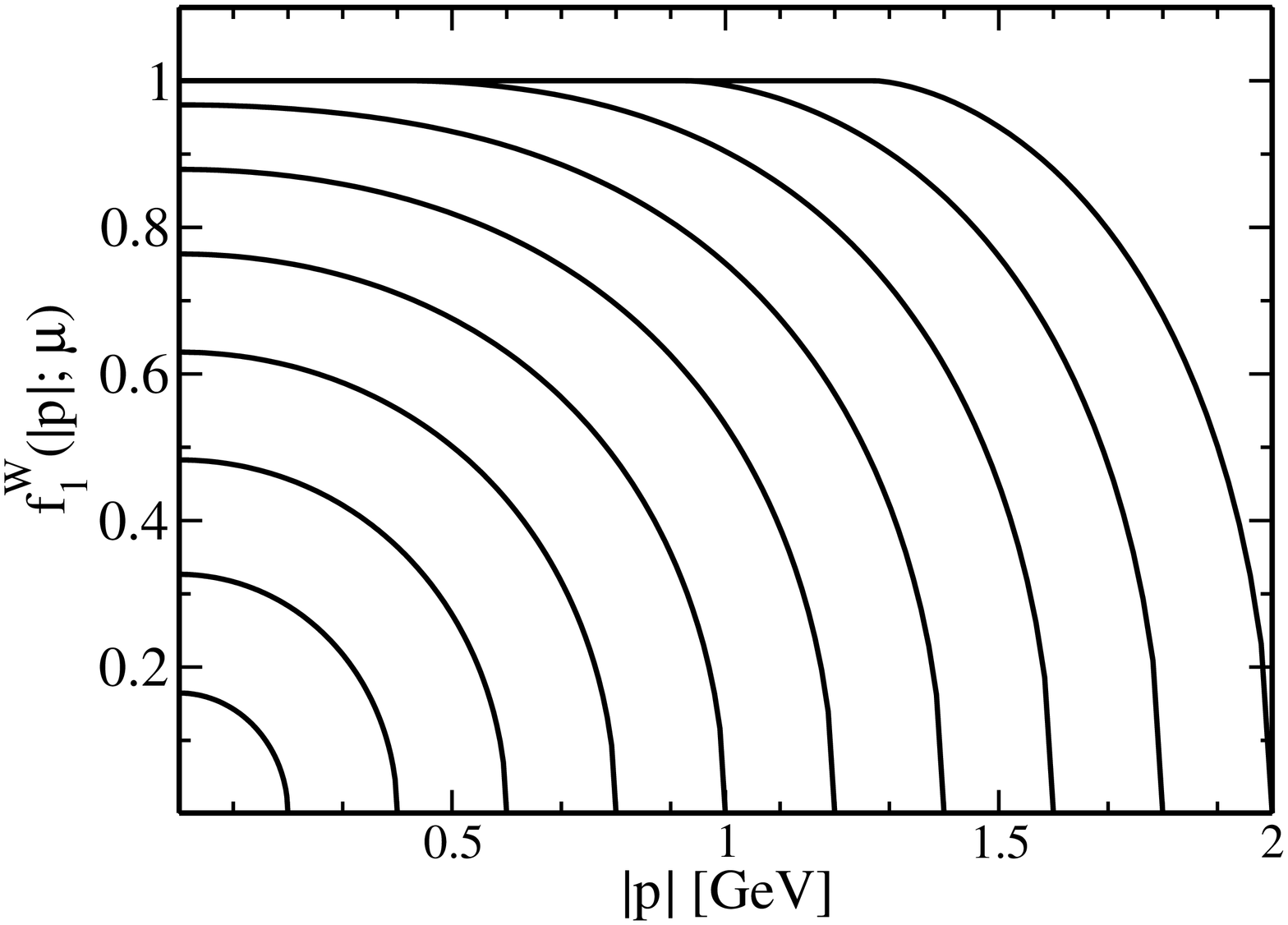}}

\caption{\label{fig:f1WI}
Wigner-phase quark number density distribution, Eqs.\,(\protect\ref{f1W}), (\ref{etavalue}).
\emph{Upper panel} --  $\mu=0.5,1,2,4,8,16\,$GeV: the Fermi gas result is recovered with increasing $\mu$.
\emph{Lower panel} --
$f_1^W(0;\mu)$ increases with increasing $\mu$ for $\mu^2<2\eta^2$, illustrated at equidistant $\mu\in[0.2,2.0]\,$GeV.  At fixed $\mu^2>2\eta^2$, $f_1$ drops from 1 to 0 on the domain $\mu^2-2\eta^2<\vec p^2<\mu^2$.
}
\end{figure}

The noninteracting Fermi gas result is recovered from Eq.\,(\ref{f1W}) when $\mu\gg\eta$; viz. [upper panel of Fig.\,\ref{fig:f1WI}],
\begin{equation}
f_1^W(\pvecabs;\mu) \stackrel{\eta/\mu\ll 1}{\approx} \theta(|\vec{p}| - \mu)\,.
\end{equation}
Novel features of the single-quark number density distribution are only exposed when $\mu \lesssim \eta$.  They originate in the dressed-quark propagator's branch point at $\vec{p}\,^2=\mu^2-2\eta^2$ [see Eq.\,(\ref{CABW})].

Whether this curious behaviour is important in the consideration of compact stars and heavy ion collisions depends on the natural scale for $\eta$.  That scale is set via the meson spectrum, a procedure which yields \cite{Munczek:1983dx}
\begin{equation}
\label{etavalue}
\eta \approx \surd 2 \, m_\rho = 1.09\,{\rm GeV}.
\end{equation}
This being so, the unconventional behaviour is important and one cannot justify the treatment of quark matter as a quasi-ideal Fermi gas; e.g., as a system in which the medium serves only to produce a density-dependent shift in the chemical potential.  Emphasis is found in the following facts.  For $0 < \mu < \surd 2 \eta$, the noninteracting Fermi gas result $f_1^W(\pvecabs;\mu)=1$ is precluded.  Indeed, it follows from Eq.\,(\ref{eq:f1WItwo}) and is illustrated in the lower panel of Fig.\,\ref{fig:f1WI}, that on this $\mu$-domain the number density distribution is $<1,\;\forall |\vec{p}|$.  Furthermore,
\begin{equation}
\label{eq:f1_p0_musmall}
f_1^W(\pvecabs=0;\mu) \stackrel{\mu^2 \ll 2\eta^2}{=}\frac{2\sqrt{2}}{\pi}\frac{\mu}{\eta};
\end{equation}
i.e., at zero momentum the number density increases linearly with chemical potential.

The Nambu-phase single-quark number density is obtained from Eqs.\,(\ref{CNG}), (\ref{BNG}), (\ref{nqmuf1}); viz.,
\begin{eqnarray}
\nonumber
4\pi \, f_1^N(\pvecabs;\mu) & = &
\int_{-p_4^*}^{p_4^*} {\rm d}p_4
\trD\left[
-\gamma_4 S_N(\pvecabs,p_4;\mu)
\right]\\
&&
\rule{-4em}{0ex}
+ 2\Real\int_{p_4^*}^{\infty}{\rm d}p_4
\trD\left[ -\gamma_4 S_W(\pvecabs,p_4;\mu) \right],
\label{eq:f1NAdef}
\end{eqnarray}
with $p_4^*=\Real\sqrt{\mu^2+\eta^2/4-\vec p\,^2}$.  Evidently, only the $2^{nd}$ term on the rhs contributes for $\vec p\,^2\ge\mu^2+\eta^2/4$ and it evaluates to the last line of Eq.\,(\ref{f1W}), so that
\begin{equation}
\label{f1NW}
f_1^N(\pvecabs;\mu) \stackrel{\vec p\,^2\ge\mu^2+\eta^2/4}{=} 0 \,.
\end{equation}

\begin{figure}[t]
\vspace*{-5ex}

\centerline{\includegraphics[width=0.33\textwidth,angle=-90,clip]{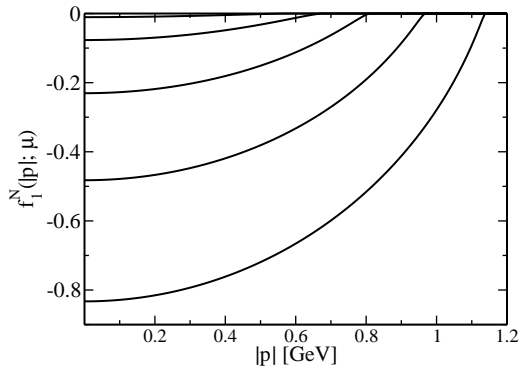}}
\caption{ \label{fig:f1NA}
Nambu-phase quark number density distribution, Eq.\,(\protect\ref{eq:f1NA}); top to bottom: $\mu=0.2,0.4,0.6,0.8,1.0$ GeV.
The density decreases with increasing $\mu$, as it must: the confined, DCSB phase is destabilised by increasing $\mu$ and hence the pressure associated with this vacuum phase must decrease.}
\end{figure}

The Nambu-phase number density evolves in a new and unusual manner on $\vec p\,^2<\mu^2+\eta^2/4$; viz.,
\begin{eqnarray}
\nonumber
\lefteqn{f_1^N(\pvecabs,\mu) =
-\frac{4\mu p_4^*}{\pi\eta^2}}\\
\nonumber
&&  + \Real\frac{i}{\pi \eta^2}
\left\{\tilde p_4^{\ast 2}-
\tilde p^{*2} \sqrt{1+\frac{2\eta^2}{\tilde p^{*2}}} \right.\\
&& - \left.
\eta^2\log
	\left[1+
	\frac{\tilde p^{*2}}{\eta^2}
		\left(1+\sqrt{1+\frac{2\eta^2}{\tilde p^{*2}}}
		\right)
	\right]
\right\},
\label{eq:f1NA}
\end{eqnarray}
where $\tilde p^{\ast \, 2} = \frac{\eta^2}{4}+2i\mu\sqrt{\mu^2-\vec{p}\,^2+\frac{\eta^2}{4}}$.  The $1^{\rm st}$ term is always negative and greater in magnitude than the positive, $2^{\rm nd}$ term.  Hence, $f_1^N(\pvecabs;\mu) \leq 0$ for $\vec p\,^2 \leq \mu^2+\eta^2/4$ [see Fig.\,\ref{fig:f1NA}].
Moreover,
$f_1^N(\pvecabs;\mu\to 0^+)\to 0^-$ and
$f_1^N(\pvecabs;\mu\to \infty)\to -\infty$.
Notably, even for $\mu \simeq 0$, the Nambu-phase number density is negative-definite for $\pvecabs < \eta/2$.
It is plain that the curious features of $f_1^N$ are again closely connected with the branch point in the quark propagator [see Eqs.\,(\ref{CNG}), (\ref{BNG})], which in this phase is intimately associated with confinement and DCSB.

The chiral-limit condensate can be obtained from the single-quark scalar density distribution
\begin{equation}
\label{eq:f2}
f_2(\pvecabs;\mu)=
\frac{1}{4\pi}
\int_{-\infty}^{\infty}{\rm d}p_4\,
{\rm tr}_{\rm D}\left\{
S(\pvecabs,p_4;\mu)
\right\}.
\end{equation}
Since the chiral-limit dressed-quark propagator is traceless in the Wigner phase, this density is identically zero.

\begin{figure}[t]
\vspace*{-5ex}

\centerline{\includegraphics[width=0.33\textwidth,angle=-90,clip]{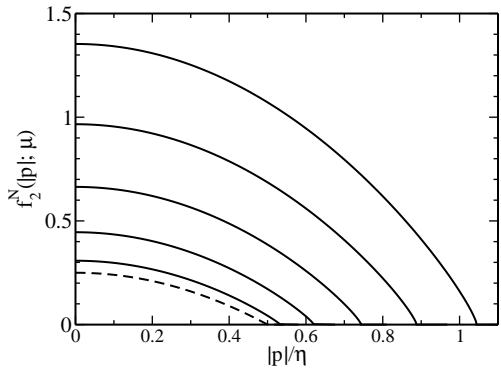}}
\caption{ \label{fig:f2NA}
Nambu-phase single-quark scalar density distribution, $f_2^N(\pvecabs;\mu)$ in Eq.\,(\protect\ref{eq:f2NA}).  Dashed curve --  $\mu=0$; and \emph{solid curves} -- $\mu=0.2,0.4,0.6,0.8,1.0$ GeV.}
\end{figure}

On the other hand, in the Nambu phase one obtains
\begin{eqnarray}
f_2^N(\pvecabs;\mu) & = & {\cal I}_{f_2}(p_4^\ast,\pvecabs;\mu) - {\cal I}_{f_2}(0,\pvecabs;\mu)\,,
\label{eq:f2NA}
\end{eqnarray}
with $(z_\mu = z+i\mu)$
\begin{eqnarray}
\nonumber
\lefteqn{{\cal I}_{f_2}(z,\pvecabs;\mu) =
\Real \left[
\frac{2 z_\mu}
{\pi\eta}\sqrt{\frac{1}{4} - \frac{\vec{p}\,^2+z_\mu^2}{\eta^2}} \right.} \\
&& \left. +
\left(\frac{1}{4} - \frac{\vec{p}\,^2}{\eta^2}\right)
\atan
\left(
  \frac{\bar z_\mu/\eta}{\sqrt{\frac{1}{4}
- \frac{\vec{p}\,^2+\bar z_\mu^2}{\eta^2}}}
\right)
\right],
\label{eq:f2NI}
\end{eqnarray}
which is positive-semi-definite, as illustrated in Fig.\,\ref{fig:f2NA}.  This expression yields
\begin{equation}
\lim_{\pvecabs\to 0;\mu\to 0}f_2^N(\pvecabs;\mu) =
\left\{
\begin{array}{ll}
\frac{1}{4}& \forall\, \eta \not=0\\
0 & \mbox{if}\; \eta=0
\end{array}
\right.,
\end{equation}
which is independent of $\eta$, so long as $\eta\neq 0$.  Since the scalar density distribution is dimensionless, an $\eta$-independent result had to be obtained in this limit.  Naturally, the scalar density distribution vanishes for $\eta = 0$; i.e., in the absence of interactions.

The chiral-limit quark condensate is
\begin{equation}
-\langle \bar q q \rangle(\mu)
=
2 N_c \int\frac{d^3\vec p}{(2\pi)^3} f_2^N(\pvecabs;\mu)
\end{equation}
and it is plain from Fig.\,\ref{fig:f2NA} that the condensate must increase in magnitude with increasing $\mu$.
Moreover, a comparison of Figs.\,\ref{fig:f1NA} and \ref{fig:f2NA} shows that the evolution of the Nambu-phase scalar density distribution is anticorrelated with that of the number distribution.  One may now understand that the decrease in the Nambu-phase pressure, which can be inferred from Fig.\,\ref{fig:f1NA}, is connected with the energy cost of rearranging the vacuum so as to increase the magnitude of the condensate in the face of opposition from the rising chemical potential.  The behaviour of hadron properties under these and similar conditions is illustrated in Refs.\,\cite{Maris:1997eg,Jiang:2008rb}.

Hitherto, general qualitative features of the pressure were used in explaining aspects of our results.  We now present this model's EoS.
As the model lies within the class of rainbow-truncations, the vacuum-pressure associated with a given phase can be calculated using the ``steepest-descent'' approximation; namely,
\begin{equation}
\label{pSigma}
P[S]=  {\rm TrLn}\left[S^{-1}\right] - \frac{1}{2}{\rm Tr}\left[\Sigma\,S\right].
\end{equation}
Equation~(\ref{pSigma}) is the auxiliary field effective action evaluated at its minimum \cite{Haymaker:1990vm}.
Owing to Eq.\,\,(\ref{Gk}), in this analysis we can neglect the gluon contribution.

\begin{figure}[t]
\vspace*{-5ex}

\centerline{\includegraphics[width=0.33\textwidth,angle=-90,clip]{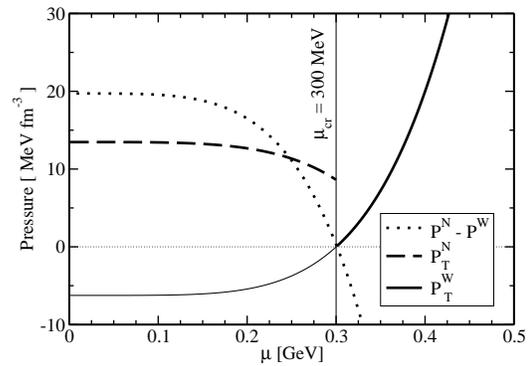}}
\caption{\label{fig:Pchiral}
\emph{Dotted curve} -- Difference between Nambu-Goldstone and Wigner-Weyl vacuum-pressures, which vanishes at $\mu_{\rm cr}=300\,$MeV;
\emph{dashed curve} -- thermodynamic pressure in the Nambu phase;
\emph{Solid curve} -- thermodynamic pressure in the Wigner phase, which is the ground state for $\mu>\mu_{\rm cr}$.}
\end{figure}

The pressure difference ${\cal B}(\mu)=P[S^N]-P[S^W]$ is computed using Eqs.\,(\ref{CABW}), (\ref{CNG}), (\ref{BNG}).
It can be identified with a bag constant \cite{Cahill:1985mh} and, for two light flavours, yields \cite{Blaschke:1997bj}
\begin{equation}
{\cal B}_{N_f=2}= (0.102\,\eta)^4=(0.111\,{\rm GeV})^4 =:\varepsilon_{\rm v}^4.
\end{equation}
As evident in Fig.\,\ref{fig:Pchiral}, a first-order transition from the Nambu-vacuum (confined, with DCSB), to the Wigner-vacuum (deconfined, chirally symmetric) occurs at
\begin{equation}
\mu_{\rm cr} := \{ \mu | {\cal B}(\mu)=0 \} = 0.276\,\eta = 300\,{\rm MeV}.
\end{equation}

The bag constant receives no contribution from $s$-quarks because their current-mass is too large to support a Wigner phase \cite{Chang:2006bm}.  This does not preclude a Cooper-paired phase involving $s$-quarks at very large $\mu$.  Also, $\varepsilon_{\rm v}$ and $\mu_{\rm cr}$ are $\sim 40$\% larger in models with an interaction whose ultraviolet behaviour is more realistic \cite{Bender:1997jf,Chen:2008zr}.

The thermodynamic pressure of each phase is determined via the quark number density.  We compute it from the generating functional in steepest-descent approximation; viz., using Eqs.\,(\ref{f1W}), (\ref{f1NW}), (\ref{eq:f1NA}) ($\phi=W,N$):
\begin{equation}
P_{\rm T}^\phi(\mu) = P_{{\rm T}}^{0\phi }+ \int_0^\mu\! {\rm d}z \, n_{\rm t}^\phi(z)\,,
n_{\rm t}^\phi(\mu) = \frac{2 N_c}{\pi^2} \int_0^\mu \! {\rm d}z z^2  f^\phi(z),
\label{eq:pressure}
\end{equation}
where $P_{\rm T}^{0N}={\cal B}-P_{\rm T}^\chi$ and $P_{\rm T}^{0W}=-P_{\rm T}^\chi$, with $P_{\rm T}^\chi$ defined so that $P_{\rm T}^{W}(\mu_c)=0$.  $P_{\rm T}^{W,N}\!(\mu)$ are depicted in Fig.\,\ref{fig:Pchiral}.

\begin{figure}[t]

\centerline{\includegraphics[width=0.33\textwidth,angle=-90,clip]{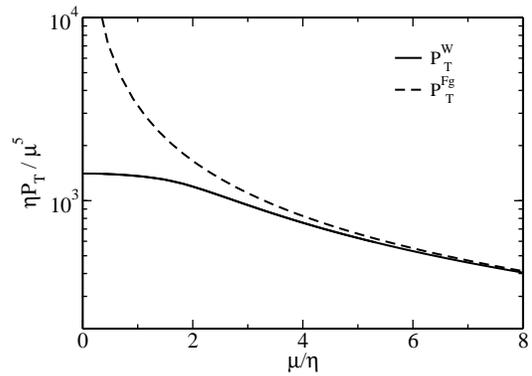}}
\caption{\label{fig:PWigner}
\emph{Solid curve} -- Thermodynamic pressure associated with the Wigner-Weyl phase of our model; and \emph{dashed curve} -- pressure of a noninteracting Fermi gas.
Evidently, $P_{\rm T}^W \propto \mu^5$ for $\mu \lesssim \eta$ and the chemical potential overwhelms the interaction for $\mu \gtrsim 4\,\eta$.}
\end{figure}

Now recall Eq.\,(\ref{eq:f1_p0_musmall}): $f_1^W(|\vec{p}|=0)\propto\mu$ for $\mu^2 \ll 2\eta^2$.  It follows, using Eqs.\,(\ref{eq:pressure}), that $P_{\rm T}^W \propto \mu^5$ on this domain.  This behaviour stands in marked contrast to that of a noninteracting Fermi gas: $P_{\rm T}^{\rm Fg} \propto \mu^4$.  The Wigner pressure is depicted in Fig.\,\ref{fig:PWigner}.

With Eqs.\,(\ref{f1W}), (\ref{f1NW}), (\ref{eq:f1NA}) and (\ref{eq:pressure}), one has the zero-temperature EoS for the rainbow-truncation of the model defined by Eq.(\ref{Gk}).  The thermodynamical energy density follows: $\varepsilon_{\rm T}(\mu)=\mu \, n(\mu)-P_{\rm T}(\mu)$.  NB.\ In assuming that $\eta$ is $\mu$-independent, we neglect quark feedback on the gluon vacuum polarisation.  In-vacuum, such effects are modest \cite{Kamleh:2007ud}.  There is currently no reason to expect otherwise in-medium.  We therefore anticipate that this feedback will have no qualitative impact on our results but may induce minor quantitative changes; e.g., a small reduction in $\mu_{\rm cr}$.

The total pressure in the confining Nambu phase receives a contribution from hadrons; viz., $P_{\rm T}^{\rm t}=P_{\rm T}^N + P_{\rm T}^H$.  However, $P_{\rm T}^H$ is omitted by the rainbow truncation, which thus precludes an internally consistent description of the transition between hadron and quark matter.  Physically, on the domain of Nambu-phase stability, one has $\partial P_{\rm T}^H/\partial \mu \geq -\partial P_{\rm T}^N/\partial \mu$; i.e., the vacuum rearrangement energy cost, evident in Fig.\,\ref{fig:f1NA}, is balanced (at least) by the gain from the response of hadron properties.  This can happen whilst maintaining $P_{\rm T}^H/P_{\rm T}^N\ll 1$.

Nevertheless, for immediate application to compact astrophysical objects, our EoS must be augmented by a model for the nuclear matter EoS, which is the active branch for small chemical potential.
Hitherto, a transition to quark matter was typically effected by a Maxwell construction and occurred when the pressures of the hadron and quark phases were equal; e.g., Refs.\,\cite{Lawley:2006ps,Blaschke:2007ri,Pagliara:2007ph,Carroll:2008sv}.
However, our understanding of the nature of the vacuum pressure obviates the need for this prescription.
Following a definition of the mapping between quark and nucleon chemical potentials, we would execute a transition to deconfined and chirally symmetric quark matter at $\mu_{\rm cr}$, with continuous pressure but discontinuous baryon number.
%

One might ask whether our quark matter EoS should be preferred over those derived from bag- or NJL-like models?  The answer is contained in the following observations. The model we have explicated is: (1) Poincar\'e covariant, symmetry preserving, exhibits both confinement and DCSB, and provides a good description of in-vacuum hadron properties; and (2) exhibits coincident deconfinement and chiral symmetry restoring transitions at nonzero temperature and chemical potential.  Neither (1) nor (2) can be said of bag- or NJL-like models.

This work was supported by:
the Department of Energy, Office of Nuclear Physics, contract no.\ DE-AC02-06CH11357;
the National Natural Science Foundation of China, contract nos.\ 10425521, 10675007, 10705002, 10735040, 10875134;
and
the Major State Basic Research Development Program, contract no.\ G2007CB815000.

\bibliography{MN_EOS}

\end{document}